\documentclass[a4paper,12pt]{scrartcl}

\usepackage[english]{babel}
\usepackage[utf8]{inputenc}
\usepackage{graphicx}
\usepackage{amsmath}
\usepackage{amssymb}

\graphicspath{{img/}}

\begin{document}

\author{V.A.~Gradusov \and S.~L.~Yakovlev\thanks{Saint Petersburg State University}}

\author{{V.\,A. Gradusov \/\thanks{e-mail: v.gradusov@spbu.ru}, S.\,L. Yakovlev\/\thanks{e-mail: s.yakovlev@spbu.ru}  }\footnote{St Petersburg State University, St Petersburg 199034, Russia}}

\title{Scattering in $e^- -(pe^-)$ and $\mu^- -(p\mu^-)$ systems: mass dependent and mass independent features of cross sections above the degenerated thresholds 
}
\date{}
\maketitle

\abstract{Ab initio calculation of low energy scattering of electrons (muons) off hydrogen (muonic hydrogen) are performed on the basis of Faddeev-Merkuriev (FM) equations. The explicit contribution of induced dipole interaction in the asymptotic behavior of the wave function components has been incorporated into FM formalism. Elastic and inelastic cross sections have been calculated with high energy resolution in the vicinity of $n=2,3$ exited states thresholds of respective atoms. The Gailitis-Dumburg oscillations are discovered in  some of calculated cross sections. }

\section{Introduction}

 We study and compare the low-energy 
 scattering of electrons $e^-$ by a hydrogen atom H(p$e^-$) and muons $\mu^-$ by a muonic hydrogen H$_{\mu^-}$(p$\mu^-$) atom in the case of zero total orbital momentum of the atom $L=0$.
Specifically we search for the above threshold pecularities in the cross sections called the Gailitis-Damburg (G-D) oscillations~\cite{Gaili63,Gaili63b,Gaili82}, which may show up  mostly for $L=0$.  
The phenomenon originates from the presence of the asymptotic dipole interaction between a charged incident (outgoing) particle and an atom that can exist in a set of degenerated Coulomb bound states.
Thus in principle it can appear in the cross sections of scattering processes in any three-body Coulomb system.
However, in reality a  form of cross sections is a complicated interplay of effects caused by both the dipole interaction and the rest part of the interaction
and therefore it is a hard task to theoretically predict the magnitudes of G-D oscillations that allow them to manifest themselves in cross sections for a specific system.
To the best of our knowledge, even in the case of the three-body Coulomb system which is fully determined 
by the values of masses and charges of its constituents, there is no such a theory that can definitely answer the question about existence of oscillations.
The influence of particle masses in a system of three charged particles have been studied in~\cite{Gaili82}.
However, the obtained statements are more related to the frequency of oscillations rather than their amplitudes.
On the other hand, measuring the low-energy electron(positron)-atom partial and full scattering cross sections in experiments is a complicated and often even impossible task.

This is the case in which the computational experiment comes to the fore  in order  to shed light on questions mentioned above.
Despite the long history of electron - hydrogen scattering calculations and its successes such as obtaining series of the below threshold narrow Feshbach resonances (see~\cite{Burke} and references therein), there are very few, if exist, high resolution in energy calculations of the right above binary threshold scattering cross sections.
One possible reason for the lack of results is the fact that these calculations are very delicate.
Actually, to solve the scattering problem in a configuration space it is critically important to use the asymptote of wave function, which explicitly takes into account the long-range particle-atom dipole potential~\cite{Hu99,Lazau18,Yak23b}.
In our work~\cite{Yak24} we have found this corrected form of the ``standard'' asymptote of scattering wave function which is very suited for scattering calculations.
We have applied it to low-energy reactive scattering in the positron-electron-antiproton system and have found the G-D oscillations in some of the obtained cross sections~\cite{Grad24}.
Our calculations are based on the theoretical and computational approach which includes solving the Faddeev-Merkuriev equations~\cite{FM, Merkur80} in the total orbital  momentum representation~\cite{Kostr89} with an efficient computational algorithm presented in~\cite{Grad21,PCT23}.
It is briefly discussed below, while the details can be found in our works~\cite{Grad21b,Grad24,Grad21}.
In this work we apply our approach to scattering of electrons $e^-$ by a hydrogen atom p$e^-$ in the case $L=0$ which looks favorable for occurrence of the G-D phenomenon~\cite{Grad24}. 
We also consider scattering of muons $\mu^-$ by a muonic hydrogen p$\mu^-$ atom which is an exotic counterpart of the $e^-$H system.

\section{Theory}
The system of three charged nonrelativistic particles with masses $m_\alpha$ and charges $Z_\alpha$, $\alpha=1,2,3$, and zero total orbital momentum $L=0$ is described by the system of Faddeev-Merkuriev equations
\begin{equation}
\label{FMeq}
\left[ T_\alpha + V_\alpha(x_\alpha)+\sum_{\beta\ne\alpha}V_\beta^{(\text{l})}(x_\beta,y_\beta) - E \right]\psi_\alpha(X_\alpha) = \\
- V_\alpha^{(\text{s})}(x_\alpha,y_\alpha)\sum_{\beta\ne\alpha}\frac{x_\alpha y_\alpha}{x_\beta y_\beta }\psi_\beta(X_\beta),
\end{equation}
$\alpha=1,2,3$.
The equations are written in the set of coordinates $X_\alpha=\{x_\alpha,y_\alpha,z_\alpha=\cos\theta_\alpha\}$ that determine the position of particles in the plane containing them. They are the lengths of the well-known reduced Jacobi vectors and the cosine of an angle between them.
The kinetic energy operators are given by
\begin{equation}
T_\alpha = -\frac{\partial^2}{\partial y_\alpha^2} -\frac{\partial^2}{\partial x_\alpha^2} - \left(\frac{1}{ y_\alpha^2}+\frac{1}{x_\alpha^2}\right)\frac{\partial}{\partial z_\alpha}(1-z_\alpha^2)\frac{\partial} {\partial z_\alpha},
\end{equation}
the potentials $V_\alpha$ represent the pair Coulomb interaction $V_{\alpha}(x_\alpha)=\sqrt{2\mu_{\alpha}}Z_\beta Z_\gamma/x_\alpha$, where $\mu_{\alpha} = m_\beta m_\gamma/(m_\beta + m_\gamma)$ is the reduced mass of pair of particles $\beta\gamma$.
They are split into short-range $V^{(\mathrm{s})}_\alpha$ and long-range parts $V^{(\mathrm{l})}_\alpha$
\begin{equation}
\label{PotSplit}
V_\alpha(x_\alpha) = V^{(\mathrm{s})}_\alpha(x_\alpha,y_\alpha) + V^{(\mathrm{l})}_\alpha(x_\alpha, y_\alpha)
\end{equation}
using a function called the Merkuriev cut-off \cite{Merkur80}.
The equations~(\ref{FMeq}) can be summed up, which leads to the Schrödinger equation for the wave function $\Psi=\sum_{\alpha}\psi_\alpha(X_\alpha)/(x_\alpha y_\alpha)$, where $\psi_{\alpha}$ are the components of the wave function.
The symmetry $p=1$ (antisymmetry $p = -1$) of the wave function with respect to a permutation of two identical particles $\alpha$ and $\beta$ is accounted for by the relations 
\begin{equation}
\psi_\beta(X_\beta) = p\psi_\alpha(-X_\beta), \quad
\psi_\gamma(X_\gamma)  =  p\psi_\gamma(-X_\gamma),
\end{equation}
where $ - X_\alpha \equiv \{x_\alpha,y_\alpha,-z_\alpha\}$.

For the scattering problem equations~(\ref{FMeq}) are complemented with zero Dirichlet boundary conditions  $x_\alpha=0$, $y_\alpha=0$ and asymptotic boundary conditions.
At energy values $E$ below the threshold of breakup (ionization) of the system, the FM components $\psi_\alpha(X_\alpha)$ at $y_\alpha \to \infty$ are substantially different from zero only in the asymptotic region $x_\alpha \ll y_\alpha $.
The corrected asymptotic conditions \cite{Yak24}  in this region take the form 
\begin{multline}
\label{asympt}
\psi_\alpha(X_\alpha)
\thicksim \sum_{n\ell}\phi_{n\ell}(x_\alpha)Y_{\ell0}(\theta_\alpha,0)
\times \bigg[ \widetilde{\psi}_{(n\ell)(n_0\ell_0)}^-(y_\alpha, p_{n_0})\delta_{\alpha\alpha_0}- \\
- \sum_{n'\ell'}\psi_{(n\ell)(n'\ell')}^+(y_\alpha, p_{n'})\sqrt{\frac{p_{n_0} }{p_{n'}}}\mathfrak{S}_{(\alpha n'\ell')(\alpha_0n_0\ell_0)} \bigg],
\end{multline}
where the incident wave is given by the expression
\begin{equation}
\left[\widetilde{\psi}^-\right]_{(n\ell)(n'\ell')}(y_\alpha,p_{n'}) \\ = \sum_{\ell''}e^ {i\left(\lambda-\lambda_{\alpha(n'\ell'')}\right)\pi/2}\left[V_{\alpha(n'\ell')(n'\ell'')}\right]^*\psi_{(n\ell)(n'\ell'')}^-(y_\alpha,p_{n'}).
\end{equation}
In these formulae the indices $\alpha n\ell$ enumerate the binary scattering channels, i.e., the Coulomb bound states of two particles $\beta\gamma$ of pair $\alpha$ with radial wave function $\phi_{n\ell}(x)$ and energy $\varepsilon_n$.
The set of indices $\alpha_0n_0\ell_0$ defines the initial scattering channel.
$Y_{\ell m}$ denotes the standard spherical harmonic.
In formula~(\ref{asympt}) and below in the text it is assumed that the indices $n\ell$ take integer values $n>\ell\ge0$ corresponding to channels which are open at a given energy $E$.
The momentum $p_n$ of an incident (scattered) particle is determined by the energy conservation condition $E = p_n^2+\varepsilon_n$.
The cross section of the scattering process with initial $\alpha_0n_0\ell_0$ and final $\alpha n\ell$ channels is expressed in a standard way through the S-matrix element
$\mathcal{S}_{(\alpha n \ell), (\alpha_0 n_0 \ell_0)}$~\cite{Grad19}, which is in turn related to the ``effective'' S-matrix $\mathfrak{S}$ (defined by the solution~(\ref{asympt})) by formula
\begin{equation}
\label{S-rec}
\mathcal{S}_{(\alpha n\ell)(\alpha_0n_0\ell_0)}
= \sum_{\ell'}e^{i\left(\ell-\lambda_{\alpha(n\ell')}\right)\pi/2}V_{\alpha(n\ell)(n\ell' )}\mathfrak{S}_{(\alpha n\ell')(\alpha_0n_0\ell_0)}.
\end{equation}
The functions $\psi_{(n\ell)(n'\ell')}^-$ and $\psi_{(n\ell)(n'\ell')}^+$ represent 
the improved incoming and outgoing waves. They are given by
\begin{equation}
\label{CC-fun}
\psi_{(n\ell)(n'\ell')}^{\pm}(y_\alpha,p_{n'})=
\left[
W^{(0)}_{\alpha(n\ell)(n'\ell')}+\frac{1}{y_\alpha^2}W^{(1)}_{\alpha(n\ell)(n'\ell')}
\right]
u_{\lambda_{\alpha(n'\ell')}}^\pm(\eta_{n'},p_{n'} y_\alpha),
\end{equation}
where $u_{\ell}^{\pm}(\eta,z)$  are the Coulomb incoming and outgoing waves~\cite{Mess},
and the Sommerfeld parameter is defined as $\eta_n\equiv Z_\alpha(Z_\beta+Z_\gamma)\sqrt{2\mu_{\alpha(\beta\gamma)}}/(2p_n)$ with $\mu_{\alpha(\beta\gamma)}=m_\alpha(m_\beta+m_\gamma)/(m_\alpha+m_\beta+m_\gamma)$ being the reduced mass of particle-atom system.
Here the matrices $W^{(0)}_\alpha$ and $W^{(1)}_\alpha$ are given by the formulae
\begin{eqnarray}
W^{(0)}_{\alpha(n\ell)(n'\ell')} & = & \delta_{nn'}V_{\alpha(n\ell)(n\ell')}, \nonumber\\
W^{(1)}_{\alpha(n\ell)(n'\ell')} & = & (1-\delta_{nn'})\frac{\sum_{\ell''=0}^ {n'-1}d_{\alpha(n\ell)(n'\ell'')}V_{\alpha(n'\ell'')(n'\ell')}}{(p_n^2-p_{n'}^2)},
\end{eqnarray}
and new values of orbital momentum $\lambda_{\alpha(n\ell)}$ are the solutions to the quadratic equation
\begin{equation}
\lambda_{\alpha(n\ell)}(\lambda_{\alpha(n\ell)}+1) = b_{\alpha(n\ell)}.
\end{equation}
Finally, $b_{\alpha(n\ell)}$ and $V_{\alpha(n\ell)}$ are the eigenvalues and eigenvectors of the matrix
\begin{equation}
\ell(\ell+1)\delta_{\ell\ell'}+d_{\alpha(n\ell)(n\ell')},\quad
\ell,\ell' = 0,1,\ldots,n-1.
\end{equation}
The elements of the matrix $d_\alpha$, which specifies the effective dipole potential, are given by the expression
\begin{equation}
\label{Ann'}
d_{\alpha(n\ell)(n'\ell')} =
D_\alpha K_{ (n\ell)(n'\ell')}
\times \bigg( \delta_{\ell',\ell+1}\frac{\ell+1}{\sqrt{4(\ell+1)^2-1}}
+ \delta_{\ell',\ell-1}\frac{\ell}{\sqrt{4\ell^2-1}} \bigg). 
\end{equation}
Here   $K_{ (n\ell)(n'\ell')} $ is  the universal integral between the radial Coulomb functions $\widetilde{\phi}_{n\ell}(x)$ with unit mass and charges
\begin{equation}
K_{(n\ell)(n'\ell')} \equiv \int_0^{+\infty}\text{d}x \widetilde{\phi}_{n'\ell'}(x)x \widetilde{\phi}_{n\ell}(x),  
\end{equation}
whereas the constant
\begin{equation}
D_\alpha = \mu_{\alpha(\beta\gamma)}\sum_{\beta\ne\alpha}(-1)^{\beta-\alpha}\text{sgn}(\beta-\alpha)\frac{Z_\alpha}{Z_\beta}\frac{1}{m_\gamma}
\end{equation}
comprises all the information about a physical system under consideration.

\section{Results}
To obtain the presented in the article results, we have calculated the scattering cross sections with a relative error of no more than 1\% and the high energy resolution: $6\cdot 10^{-6}$ a.u. when calculating cross sections for the e${}^-$-H system and $1.2\cdot 10^{-4}$ a.u. for the $\mu^-$-$\text{H}_{\mu^-}$ system.
All presented values are given in atomic units, cross sections are given
in units of $\pi a_0^2$.
Binary scattering processes are specified by the initial and final atomic states. For example, ${\mathrm{H}(2p)\to\mathrm{H}(3d)}$ denotes a hydrogen excitation process with $n=2$ $p$ initial state and $n=3$ $d$ final state. All  cross sections are given for singlet states with no spin weight included. The doublet cross sections will be presented in a forthcoming publication.  

According to the G-D theory~\cite{Gaili82}, the near-threshold oscillations in cross sections arise when some of the new values of orbital momentum $\lambda_{\alpha(n\ell)}$ are complex.
Above the threshold of the excited bound state of an atom with principal quantum number $n$, in the case of a single (among values with different $\ell<n$) non real value $\lambda_{\alpha(n\ell)}$, the theory predicts the following dependence of the cross sections on energies $p_n^2$:
\begin{equation}
\label{GD}
\sigma =
A + B\cos(\Im\mbox{m} \,\lambda_{\alpha(n\ell)}\ln p^2_n+\phi).
\end{equation}
Here the constants $A$, $B$, $\phi$, their own for each specific system and cross section, can be considered to be independent of the energy $p_n^2$ for small $p_n$.
A simple calculation shows that nonzero $\Im\mbox{m} \,\lambda_{\alpha(n\ell)}$ for $n=2$ and $n=3$ equals 2.198 and 3.994, correspondingly, for the $e^-$-H and 2.212 and 4.017 for the $\mu^-$-$\text{H}_{\mu^-}$ system case.
That is, the frequency values of the G-D oscillations described by~(\ref{GD}) are very close for these two physical systems.
This is not surprising since the constant $D_\alpha$ equals -1.000000296 and -1.0103 for $e^-$-H and $\mu^-$-$\text{H}_{\mu^-}$ system cases, correspondingly.

\begin{figure}[t]
	
	\begin{minipage}[h]{0.49\textwidth}
		\center{\includegraphics[width=1\textwidth]{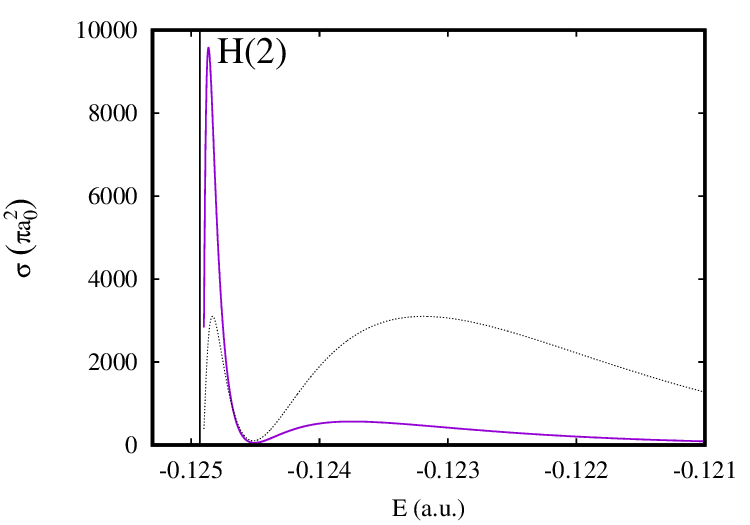}} \\ (a)
	\end{minipage}
	\hfill
	\begin{minipage}[h]{0.49\textwidth}
		\center{\includegraphics[width=1\textwidth]{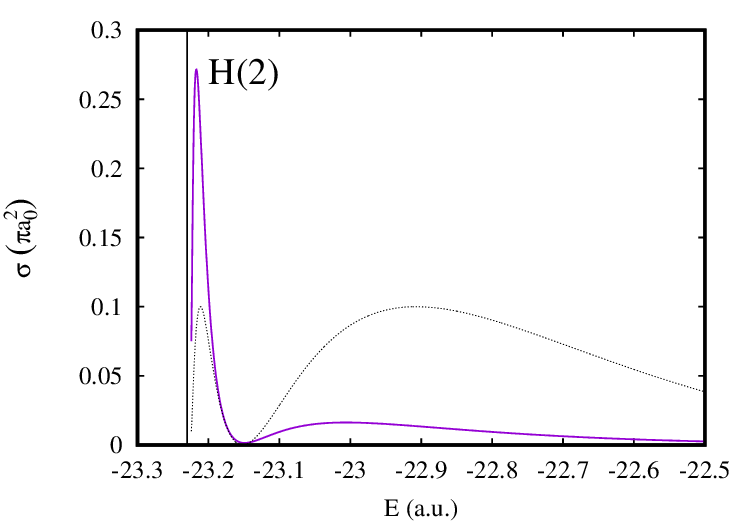}} \\ (b)
	\end{minipage}
	
	\begin{minipage}[h]{0.49\textwidth}
		\center{\includegraphics[width=1\textwidth]{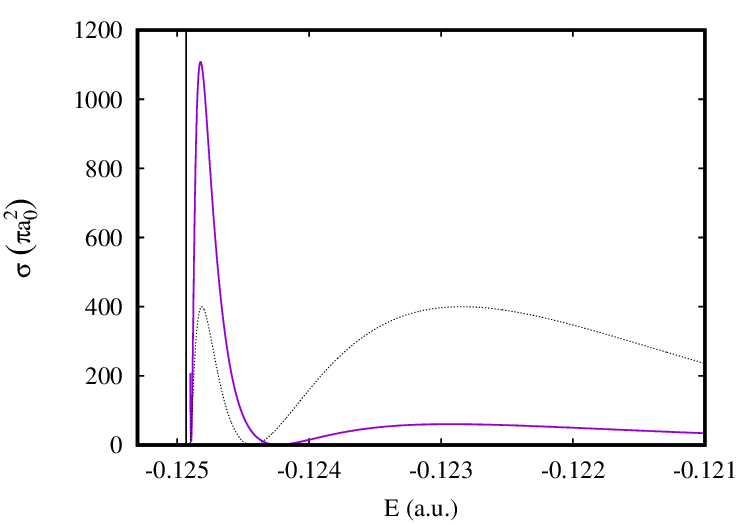}} \\ (c)
	\end{minipage}
	\hfill
	\begin{minipage}[h]{0.49\textwidth}
		\center{\includegraphics[width=1\textwidth]{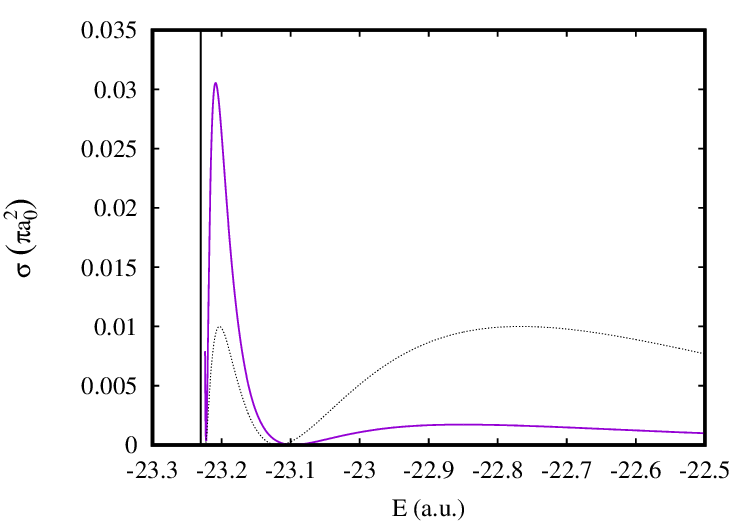}} \\ (d)
	\end{minipage}
	
	\caption{Cross sections of: elastic (a) H(2s)$\to$H(2s) and (b) $\text{H}_\mu$(2s)$\to$ $\text{H}_\mu$(2s), quasielastic  (c) H(2p)$\to$H(2s) and (d) $\text{H}_\mu$(2p)$\to$ $\text{H}_\mu$(2s) scattering processes. The dotted line shows a possible form of the curve~(\ref{GD}). }
	\label{above_H2}
	
\end{figure}

Figure~\ref{above_H2} shows elastic and quasielastic cross sections for $e^-$ ($\mu^-$) scattering off the first excited state of H ($\text{H}_{\mu^-}$) at energies above this state threshold.
The clearly visible peaks and basins can be identified as G-D oscillations since their locations are in good agreement with the law~(\ref{GD}).
Indeed, for clarity, Figure~\ref{above_H2} also shows curves~(\ref{GD}) with empirically chosen values of constants $A$, $B$ and phase $\phi$.

\begin{figure}[t]
	
	\begin{minipage}[h]{0.49\textwidth}
		\center{\includegraphics[width=1\textwidth]{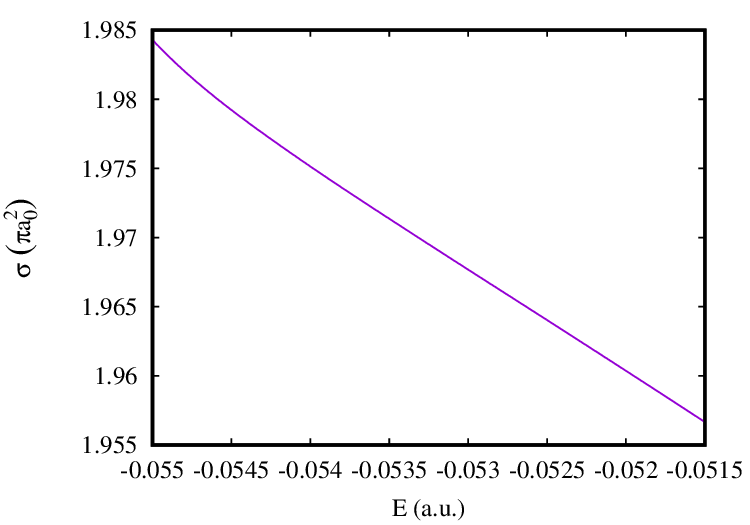}} \\ (a)
	\end{minipage}
	\hfill
	\begin{minipage}[h]{0.49\textwidth}
		\center{\includegraphics[width=1\textwidth]{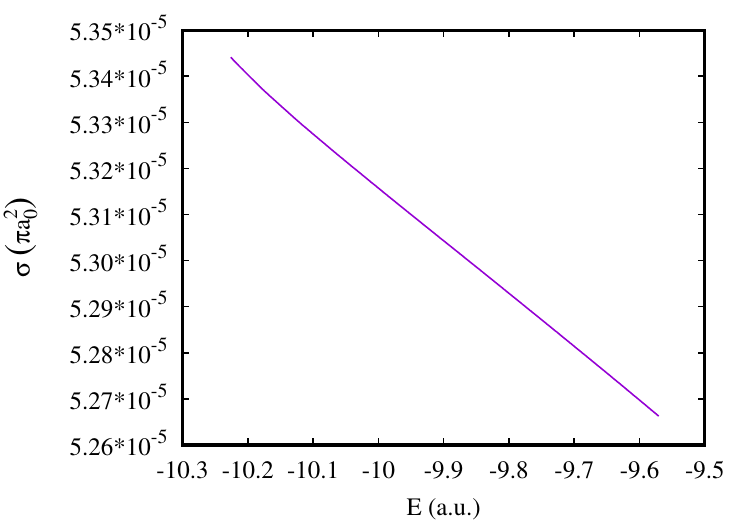}} \\ (b)
	\end{minipage}
	
	\begin{minipage}[h]{0.49\textwidth}
		\center{\includegraphics[width=1\textwidth]{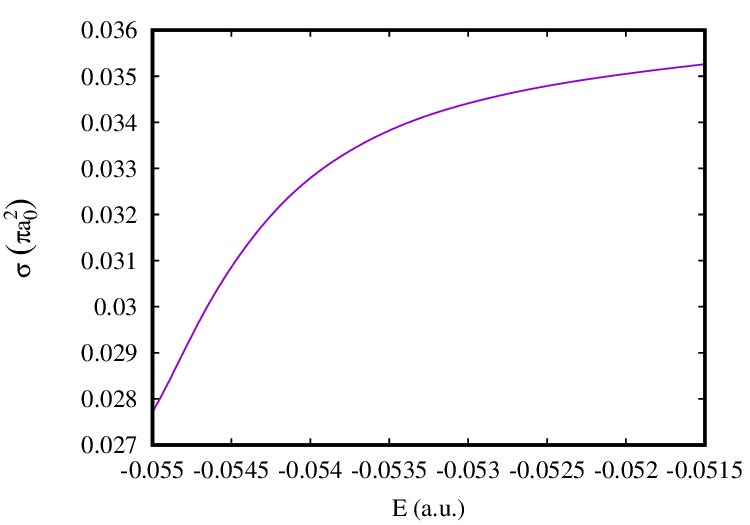}} \\ (c)
	\end{minipage}
	\hfill
	\begin{minipage}[h]{0.49\textwidth}
		\center{\includegraphics[width=1\textwidth]{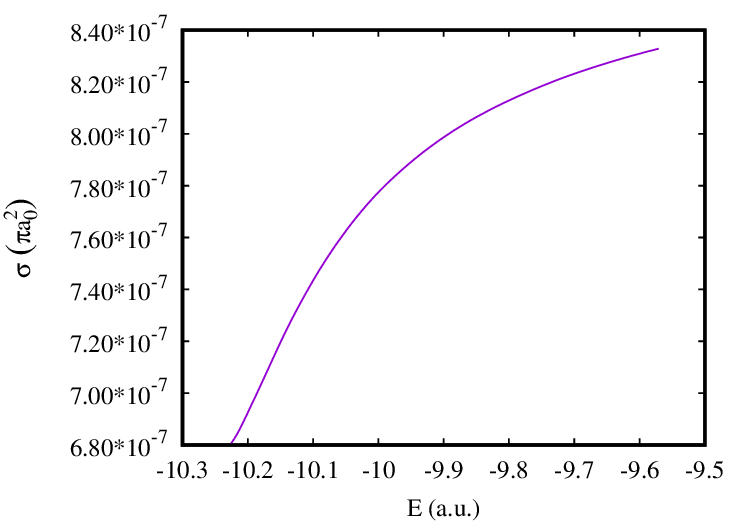}} \\ (d)
	\end{minipage}
	
	\begin{minipage}[h]{0.49\textwidth}
		\center{\includegraphics[width=1\textwidth]{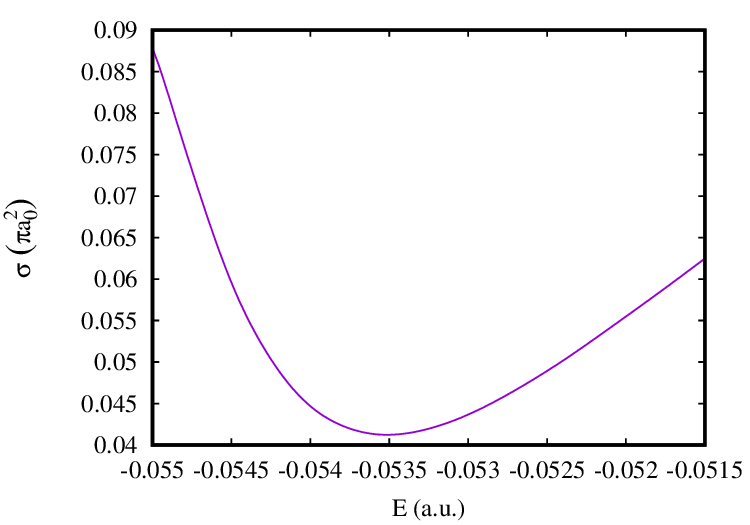}} \\ (e)
	\end{minipage}
	\hfill
	\begin{minipage}[h]{0.49\textwidth}
		\center{\includegraphics[width=1\textwidth]{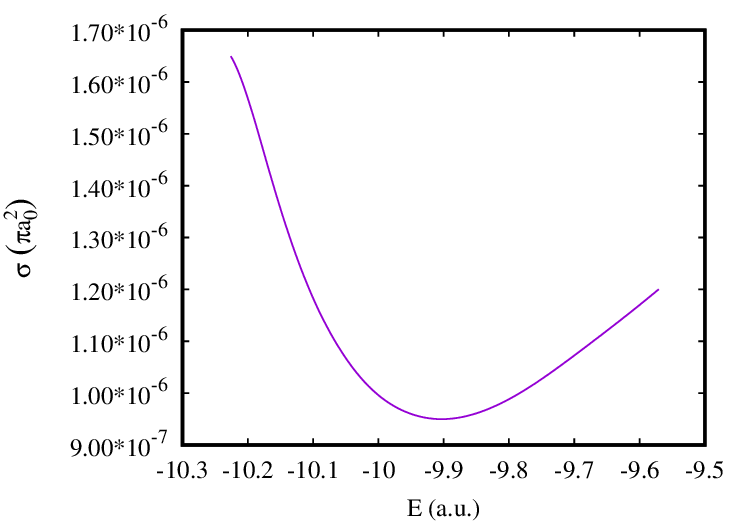}} \\ (f)
	\end{minipage}
	
	\caption{Cross sections of: (a) H(1s)$\to$H(1s), (b) $\text{H}_\mu$(1s)$\to$ $\text{H}_\mu$(1s), (c) H(1s)$\to$H(3p), (d) $\text{H}_\mu$(1s)$\to$ $\text{H}_\mu$(3p), (e) H(2p)$\to$H(3d), (f) $\text{H}_\mu$(2p)$\to$ $\text{H}_\mu$(3d) scattering processes.}
	\label{above_H3}
	
\end{figure}

\begin{figure}[t]
	
	\begin{minipage}[h]{0.49\textwidth}
		\center{\includegraphics[width=1\textwidth]{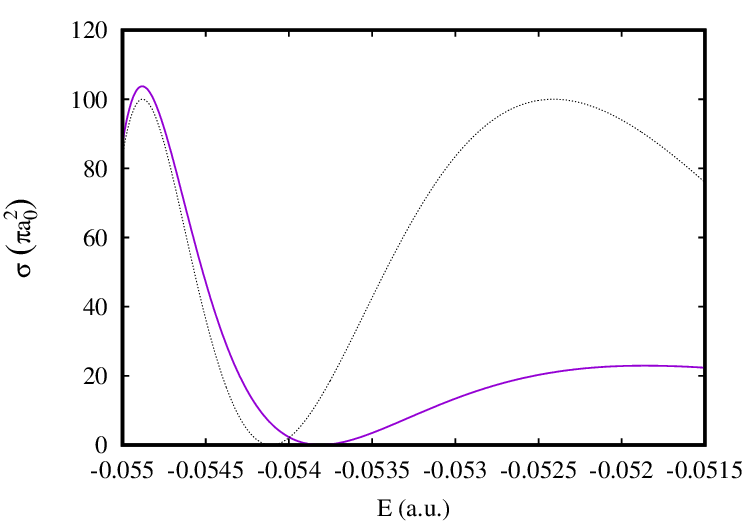}} \\ (a)
	\end{minipage}
	\hfill
	\begin{minipage}[h]{0.49\textwidth}
		\center{\includegraphics[width=1\textwidth]{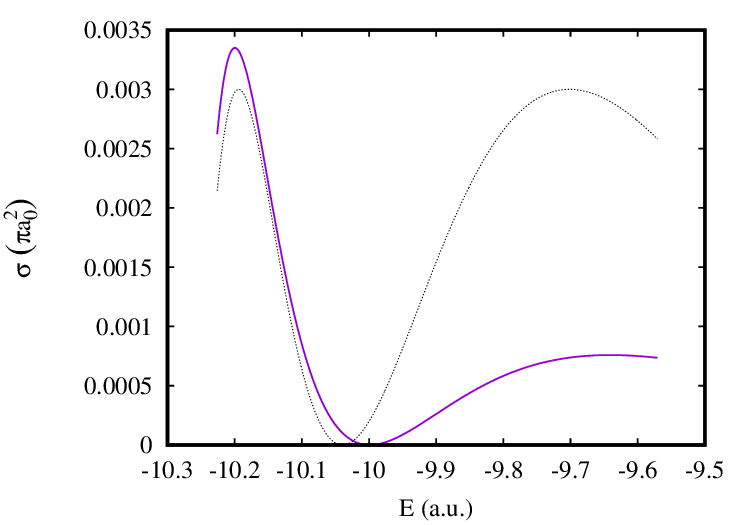}} \\ (b)
	\end{minipage}
	
	\begin{minipage}[h]{0.49\textwidth}
		\center{\includegraphics[width=1\textwidth]{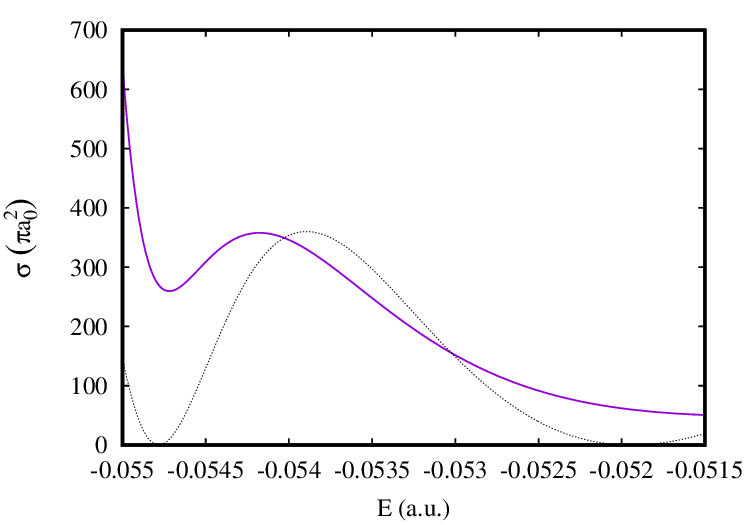}} \\ (c)
	\end{minipage}
	\hfill
	\begin{minipage}[h]{0.49\textwidth}
		\center{\includegraphics[width=1\textwidth]{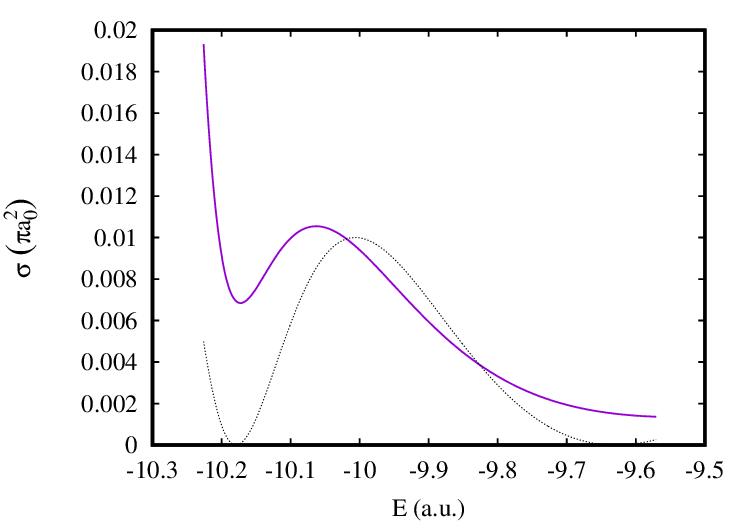}} \\ (d)
	\end{minipage}
	
	\begin{minipage}[h]{0.49\textwidth}
		\center{\includegraphics[width=1\textwidth]{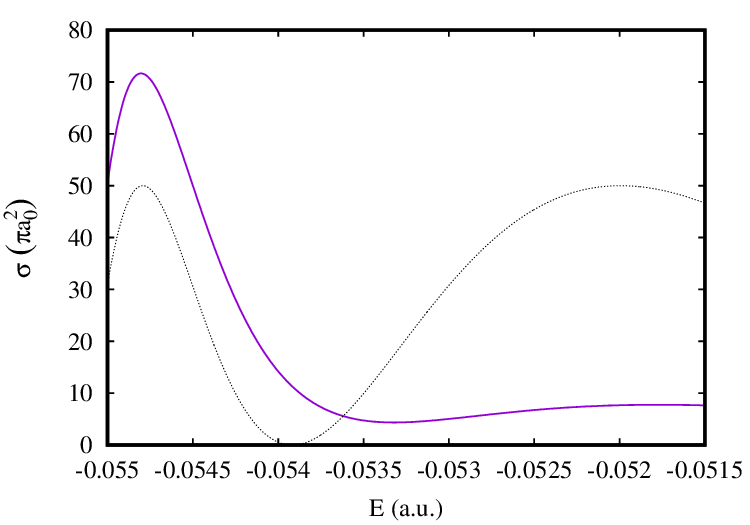}} \\ (e)
	\end{minipage}
	\hfill
	\begin{minipage}[h]{0.49\textwidth}
		\center{\includegraphics[width=1\textwidth]{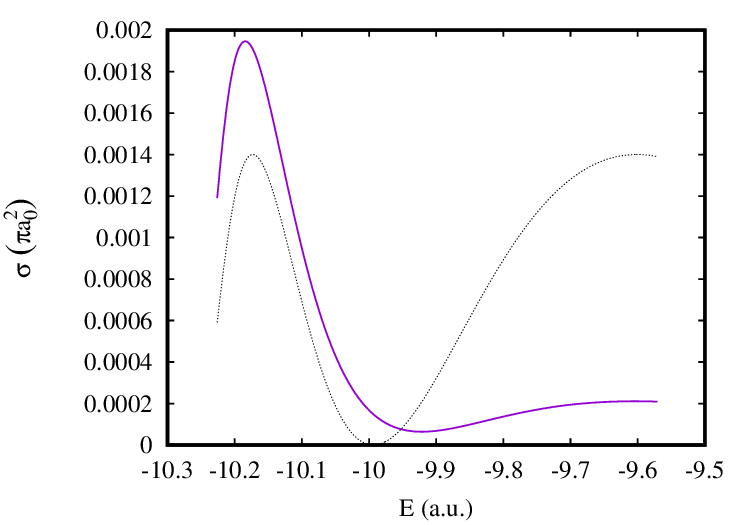}} \\ (f)
	\end{minipage}
	
	\caption{Cross sections of: (a) H(3p)$\to$H(3s), (b) $\text{H}_\mu$(3p)$\to$ $\text{H}_\mu$(3s), (c) H(3p)$\to$H(3p), (d) $\text{H}_\mu$(3p)$\to$ $\text{H}_\mu$(3p), (e) H(3p)$\to$H(3d), (f) $\text{H}_\mu$(3p)$\to$ $\text{H}_\mu$(3d) scattering processes. The dotted line shows a possible form of the curve~(\ref{GD}). }
	\label{above_H3_GD}
	
\end{figure}

Figures~\ref{above_H3} and~\ref{above_H3_GD} show cross sections of various elastic and inelastic $e^-$-H and $\mu^-$-$\text{H}_{\mu^-}$ scattering processes above the second excited state H(3) and  $\text{H}_{\mu^-}$(3) thresholds respectively.
Some of the cross sections which are presented in Figure~\ref{above_H3_GD} show the signs of G-D oscillations while other presented in Figure~\ref{above_H3} do not.

\begin{figure}[t]
	
	\begin{minipage}[h]{0.49\textwidth}
		\center{\includegraphics[width=1\textwidth]{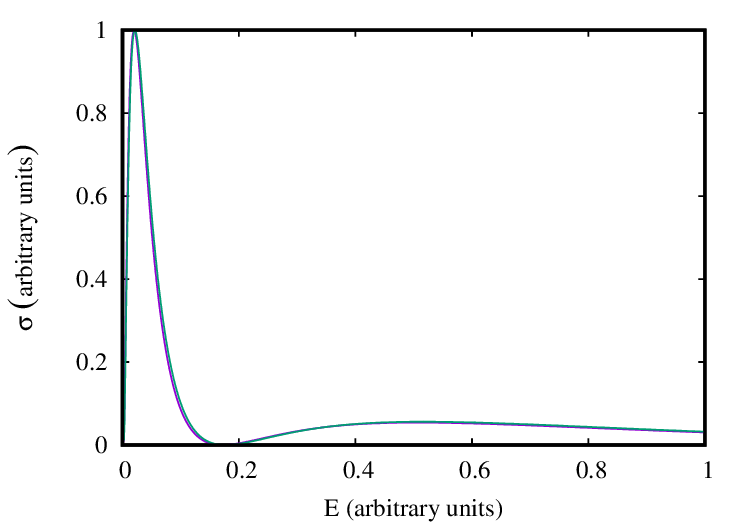}} \\ (a)
	\end{minipage}
	\hfill
	\begin{minipage}[h]{0.49\textwidth}
		\center{\includegraphics[width=1\textwidth]{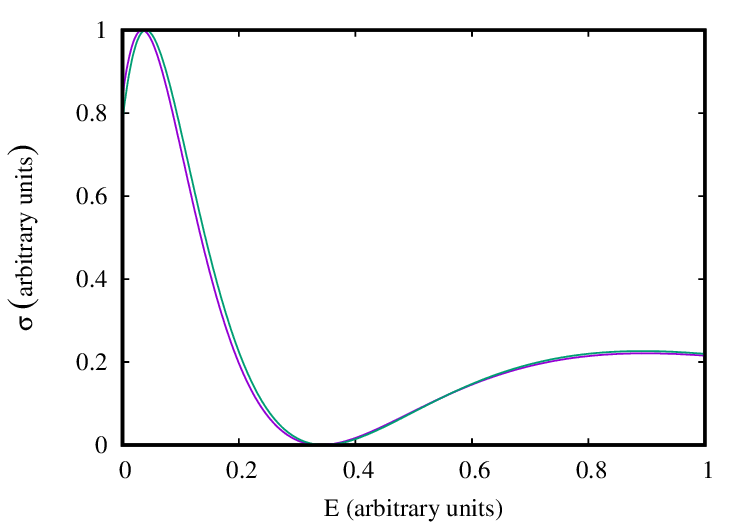}} \\ (b)
	\end{minipage}
	
	\begin{minipage}[h]{0.49\textwidth}
		\center{\includegraphics[width=1\textwidth]{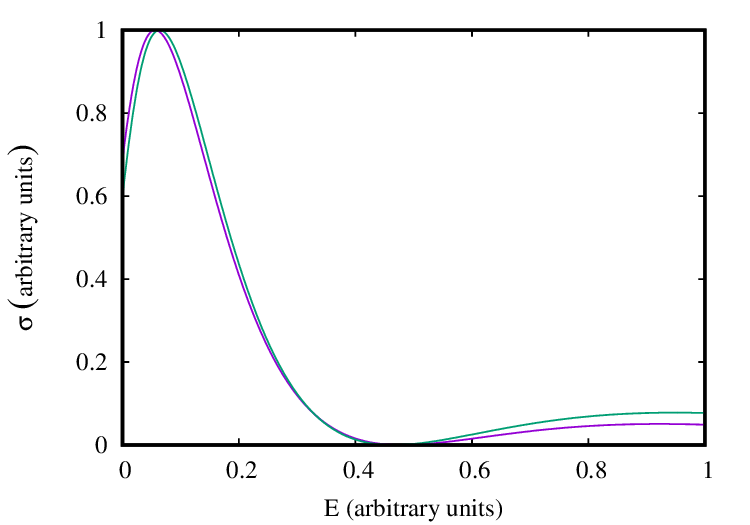}} \\ (c)
	\end{minipage}
	\hfill
	\begin{minipage}[h]{0.49\textwidth}
		\center{\includegraphics[width=1\textwidth]{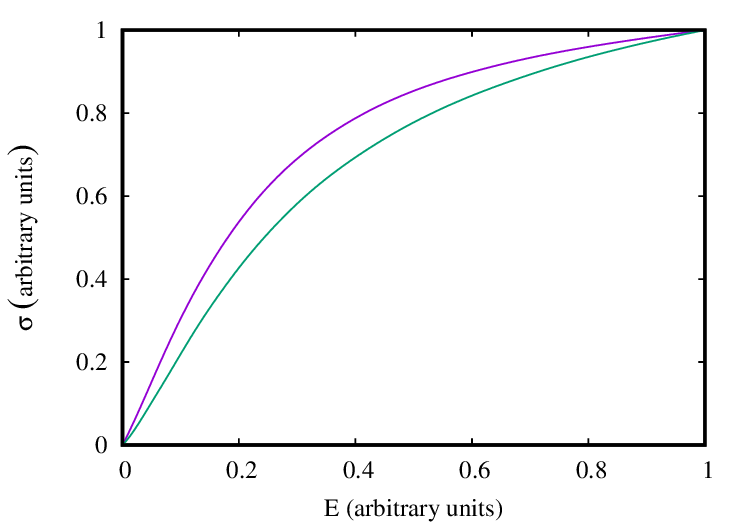}} \\ (d)
	\end{minipage}
	
	\caption{Comparison of cross sections: (a) H(2p)$\to$H(2s) and $\text{H}_\mu$(2p)$\to$ $\text{H}_\mu$(2s), (b) H(3p)$\to$H(3s) and $\text{H}_\mu$(3p)$\to$ $\text{H}_\mu$(3s), (c) H(3p)$\to$H(3d) and $\text{H}_\mu$(3p)$\to$ $\text{H}_\mu$(3d), (d) H(1s)$\to$H(3p) and $\text{H}_\mu$(1s)$\to$ $\text{H}_\mu$(3p). Purple lines correspond to $e^-$-H, green lines to $\mu^-$-$\text{H}_{\mu^-}$ scattering cross sections.}
	\label{evsmu}
	
\end{figure}

As it is seen from the presented figures, although the energy and cross section value scales for $e^-$-H and $\mu^-$-$\text{H}_{\mu^-}$ systems differ significantly, the cross section curve forms are very similar. To make this assertion more convincingly we have replotted in Figure~\ref{evsmu} some of the presented on previous plots cross sections placing the curves corresponding to similar processes on one graph.
On this graph all the $x$ and $y$ axes intervals are rescaled linearly to fit each other.
The interesting observation is that those cross sections in which the G-D oscillations are present, i.e. (a), (b) and (c) curves, are almost identical for the considered systems.
It can imply the domination of the G-D mechanism in the corresponding scattering processes and universality of the law~(\ref{GD}) for systems with identical particle-atom dipole potentials (see the discussion about the constant $D_\alpha$ above). 

\section{Summary}
We have calculated various $L=0$ cross sections of scattering in $e^-$-H and $\mu^-$-$\text{H}_{\mu^-}$ systems above the first two thresholds of excited  states of respective atoms and discovered the G-D oscillations in some of them.
These above threshold cross sections are very similar for the two physical systems considered.
We have discussed the possible origin of this similarity.

\section*{Acknowledgment} 
Research was carried out using computational resources
provided by Resource Center “Computer Center of SPbU”
(http://cc.spbu.ru)

\section*{Funding}
This work was supported by the Russian Science Foundation, project no. 23-22-00109.

\bibliography{refs}
\end{document}